# Effect of Organoclay on Compatibilization, Thermal and Mechanical Properties of Polycarbonate/Polystyrene Blends

A K Singh, Rajiv Prakash*

School of Materials Science & Technology

Indian Institute of Technology (Banaras Hindu University), Varanasi-221005, India

## Abstract

Pristine and organoclay modified polycarbonate/polystyrene (PC/PS) blends are prepared using melt-mixing technique. These blends are characterized for their morphology, structural, thermal and mechanical properties. Though our FTIR and XRD results show weak interactions between PC and PS phases, however, DSC and morphological study reveals that pristine PC/PS blends are immiscible. On other hand, introduction of organoclay results compatibilization of two polymer phases which is supported by significant shift in glass transition temperatures of the component phases and a distinct morphology having no phase segregation on sub-micron scale. Intercalation of polymers inside the clay gallery is achieved and is supported by XRD studies. A better thermal stability and higher value of modulus of the compatibilized blends compared to pristine PC/PS blends also support the reinforcement effect of organoclay to the PC/PS blend matrix.

*Author to whom correspondence should be addressed; electronic mail: rajivprakash12@yahoo.com

**1. Introduction**

Development of new polymeric materials with improved and excellent properties is in great need for many commercial applications. Polymer blending has emerged as a cheaper and easier route to develop such polymeric materials. However, because of the large unfavorable enthalpy of mixing, most polymer blends tend to macro-phase separation which results in deterioration in properties. To get the optimum blend property and long-term stability, it is required to prevent phase segregation and to stabilize the morphology of immiscible blends through suitable compatibilization techniques [1-4]. Compatibilization of two polymer phases can be achieved by adding a block, graft, or cross-linked copolymer of two polymer components in the blend, or by forming such a copolymer through covalent or ionic bond formation *in-situ*, which is commonly known as reactive compatibilization [5]. In general the added copolymers are compatible with both phases; therefore, they segregate at interface preferentially ensuring strong interfacial adhesion. But the synthesis of such copolymers with a special structure is a complex process, in many cases, which makes them quite expensive. Moreover, this expensive exercise yields only a little benefit in improving the mechanical properties [2].

Recently, a newly explored compatibilization method makes use of functionalized inorganic solid particles as compatibilizer [6-11]. This give rise to the strong adsorption of polymers on the solid surfaces and as a result the stabilizing energy gain is originated. Essentially the functionalized inorganic solid particles should have a large surface area so that the gain in stabilizing energy could overcome to the positive value of Gibb's free energy of immiscible polymer pairs. This can be achieved by using organically modified clays (organoclays) or other nanostructures having large specific surface area (such as functionalized carbon nanotubes) and able to be dispersed within a two-phase matrix [6-11].

The present work explores various aspects of PC/PS blends. The reason for choosing PC/PS blend system for this study is the surge of interest in polycarbonate (PC) because of its high transparency, high impact resistance and good heat resistance etc. However, due to its high melt viscosity and sensitivity of impact strength on temperature (especially for thick product), its processing is difficult. On the other hand, polystyrene (PS), an important engineering thermoplastic, can be easily injected, extruded/blow moulded which makes it a very useful and versatile manufacturing material. But, the most important weakness of PS is the rapid formation of crazes at high impact and deformation rates which latter on results into the formation of cracks. The main reason underlying the craze formation is the restricted movement of the molecular chains owing to large side groups. Since the properties of PC and PS are complementary to each other it is expected that PC/PS blends may overcome these drawbacks.

In PC/PS blend system, blends components are highly immiscible [12-13]. Scott [14], Kim and Burns [15] calculated the polymer-polymer interaction parameter of PC/PS blends and concluded that PC/PS blends were partially miscible blends. Further, Chaui et al. [16] and Groeninckx et al. [17], based on their morphology studies; and Kunori and Geil [18], based on their dynamic mechanical and thermomechanical studies, suggested that PC/PS blends possess very low mutual solubility which leads to the incompatible blend formation. Then after, numerous attempts have been made to enhance the compatibility of PC/PS blends. For example, Pu et al. [19] prepared PC-g-PS graft copolymer by irradiation and studied the various aspects of PC-g-Ps graft copolymer and

PC/PS/PC-g-PS ternary blends. Lee and Park [20] prepared polycarbonates with anhydride functionality and blended it with polystyrene containing oxazoline functional groups. Increase in torque of the reactive blend and signature of inter-polymer reaction were marked through experiments. Moreover, the morphology of reactive blends was found to be suppressed phase coalescence fairly compared with that of nonreactive blend. Later on they shows that polybutylene terephthalate (PBT)/oxazoline containing polystyrene graft copolymer works as a compatibilizer for PC/PS blends [21].

The foregoing discussion suggests that many of efforts for compatibilization of PC/PS blends were focused on the basis of synthesis of the copolymer which can emulsify the interfaces of the blends [19-22]. But, because of the difference in polymerization mechanism of PC and PS, the synthesis of PC/PS copolymer is difficult and complicated. To overcome these difficulties in present work we have used organoclay as nanofiller in the PC/PS blends and studied the compatibility, thermal and mechanical properties of pristine and organoclay modified blends. A detailed experimental study of the compatibility, thermal and mechanical properties of pristine and organoclay modified blends concludes that the organoclay suppresses phase separation and increases the compatibility in the blends.

## 2. Experimental

PC and PS used in this study were the commercial grade polymers. PS was purchased from Aldrich and according to supplier it has mol. weight (Mw) 192000. PC (Makrolon 2015) was purchased from Bayer; according to supplier it has mol. weight (Mn) 15000. Cloisite® 15A (with organic modifier dimethyl, dihydrogenated tallow, quaternary ammonium; cation exchange capacity 125 meq/100g clay; $d_{(001)}$=3.15 nm), were purchased from Southern Clay Product Inc. and used as received.

Initially, the PC and PS beads were converted into powder form by using a grinder. A requisite amount of PC, PS powder and organoclays; in case of blend nanocomposite were mixed well in a high-speed mixer before being put it into the extruder. Adsorbed water was removed by keeping the as mixed powder in a vacuum oven for 24h at 80°C. PC/PS blends and its blend nanocomposite in different weight ratios were made by melt extrusion method using a co-rotating twin screw extruder (Thermo Scientific Haake Minilab II). The mixing was done at a temperature of 220°C for 10 min under a shear rate of 75 rpm for the optimum blend formation. Tensile test specimens were prepared by microinjection using microinjector (Theromo Scientific, Haake Minijet II) and a mold tensile bar (557-2298, ISO 527-2-5A). All the samples were microinjected at a barrel temperature of 260°C and mold temperature of 80°C with a pressure of 700 bar.

Differential scanning calorimetric study on PC/PS blends without and with organoclays was carried out using DSC (Mettler-Toledo, 823) at a heating rate of 20°C/ minute under nitrogen atmosphere. The glass transition temperature was taken from second run after removing all the prior thermal history in first run. For this, first of all we heat the sample from 30 to 275°C at a heating rate of 20°C/min, and then cool the sample with the maximum possible cooling rate of 45°C /minute available with the DSC. This constitutes the first DSC run. After this, we reheat the sample from 30 to 275°C at a heating rate of 20°C / min for recording the DSC data in the second run. The DSC was calibrated with indium and zinc before use.

The thermal degradation was studied using TGA (Mettler-Toledo, TGA/DSC-1) at a heating rate of 20°C/ min. under nitrogen atmosphere. The degradation temperature of the samples was measured on the basis of 10% weight loss.

X-ray diffraction experiments were performed using a Bruker AXS D8 Advance wide-angle X-ray diffractometer with Cu Kα radiation and a graphite monochromator (wavelength, $\lambda$ = 0.154 nm). The generator was operated at 40 kV and 40 mA. The film samples (except for ornganoclay which was placed in powder form) were placed on a quartz sample holder at room temperature and were scanned at diffraction angle $2\theta$ from 1° to 40° at the scanning rate of 1°/ min to explore the nanostructure and effect of organoclay on blend matrix.

FT-IR spectroscopic analyses of the compression molded samples were conducted in ATR mode in a Thermo Scientific FTIR spectrometer (model: NICOLET 6700) from 650 to 4000 $cm^{-1}$ with a resolution of 2 $cm^{-1}$.

Tensile tests were carried out with the injection-molded tensile specimens using an Instron 3369 tensile tester at a strain rate of 2 mm/min at room temperature. Several samples were tested to obtain better error estimation.

The blend morphology were examined by SEM using a ZEISS (model: SEM-Supra 40) apparatus operating at an accelerating voltage of 10kV. The sample fractured in liquid nitrogen and then sputter coated with gold to avoid charging.

**3. Results and Discussion**

**3.1 Evidence for immiscible blend formation in pristine PC/PS blends**

Fig. 1 shows second run DSC thermograms of pure PC, pure PS and melt extruded PC/PS blends. Pure PS and PC show their glass transition temperature at 105 and 148°C, respectively. It is observed that all the blends composition exhibit two $T_g$s which is an indicative of immiscible blends formation. The lower $T_g$ is corresponds to the PS-rich while higher $T_g$ corresponds to the PC-rich phase of PC/PS blends. In the case of PC/30PS blends, both the $T_g$s are somewhat shifted towards higher and lower temperature sides approaching each other, which indeed suggesting weak interaction developing between PC and PS in the melt extruded blends.

Fig. 2 shows SEM images of cryo-fractured surface of PC/30PS blend. The circular region in the SEM images corresponds to the PS phase, while the continuous matrix corresponds to PC phase. The minority PS phase is dispersed all over the PC matrix with the phase segregation on sub-micron scale. This confirms the immiscibility in pristine PC/PS blends.

**3.2 Interactions between PC and PS phases**

Fig.3 shows the FTIR spectra of pure PC, PS and pristine PC/PS blends in the wave-number range 650-2000 cm$^{-1}$. In the case of blend system, IR peak (750 cm$^{-1}$) corresponding to wagging motion of five hydrogen atoms and the stretching vibration peaks (1492, 1600 cm$^{-1}$) corresponding to benzene ring of polystyrene and the starching vibration peak corresponding to polycarbonate (1500 cm$^{-1}$) slightly shifted. This shift in peak position of different bands along with a small change in their intensity may attribute to the interaction between PC and PS phases of the blends.

Fig. 4 shows XRD patterns of the melt extruded PC, PS and PC/PS blends. All the XRD patterns are diffuse in shape which is an indicative of amorphous nature of polymers. Pure PC shows a single diffuse XRD peak centered at 2θ~17.2° which is known to be due to the interference between the chains (Mitchell and Windle [23];

Windle [24]). Similarly, pure PS also shows a broad peak centered at 2θ~ 19.3°. In the PC/PS blend, as the PS content increases, a significant shift in the characteristic XRD peak of PC (at 2θ ~17.2°) towards higher two-theta side is observed. Inset to the Fig. 4 shows a variation of the peak position of the characteristic peak of PC with a variation in the content of PS in PC/PS blends. This again suggests an interaction between PC and PS phases in melt extruded PC/PS blends [25].

**3.3 Effect of organoclay on compatibility of PC/PS blends**

Cloisite® 15A (C15A), the organoclay, has been added in the blend to make the nanocomposite. It has slightly higher degradation temperature 275°C (measured at 5% weight loss, TGA were carried out at a heating rate of 20°C/min. under nitrogen atmosphere) than other organoclays (for example degradation temperature corresponding to 5% weight loss is found to be 222°C, 272°C for C10A and C30B organoclays, respectively). The PC/PS blends with 3%C15A organoclay were prepared by melt extrusion technique and the effect of organoclay on the compatibility, thermal and mechanical properties, etc were studied.

Fig. 5 shows second run DSC thermograms for all these samples. As discussed earlier, pristine PC/PS shows two distinct $T_g$ corresponding to PS and PC phases. Also the shift in $T_g$s is insignificant with respect to $T_g$s of pure PS and PC. In the organoclay modified nanocomposite/blend a comparatively large shift in $T_g$s of the PS and PC are observed. The $T_g$s for C15A modified PC/10PS and PC/30PS blends are (106°C, 139°C) and (106°C, 132°C), respectively. This significant shift in $T_g$s may attribute to the compatibilization of PC and PS phases of the blends in presence of C15A organoclay. This compatibilization may result because the organoclay has better interaction with PS and PC than PC/PS itself, which lowers the Gibb's free energy of mixing and increase the miscibility of the component phases. Also the organoclay platelets may act like a "knife" thereby reducing dispersed phase domain size due to shear stress generated during compounding of the blends and improving the compatibility of the blends.

Fig. 6 shows morphology of organoclay modified PC/30PS blends. A distinct morphology is seen in the SEM of organoclay modified PC/30PS blends than observed for pristine PC/30PS (shown in Fig.2). We did not observe any phase segregated region in the morphology. This suggests the compatibilization of PC and PS phases of the blends in presence of organoclay.

### 3.4 Nanostructure of organoclay modified blends

To understand the state of dispersion of organoclay in PC/PS blends, XRD analysis of C15A powder and the organoclay modified PC and PC/30PS blends were carried out (shown in Fig. 7). The characteristic XRD peaks of C15A organoclay, which represents the interplanar distance of silicate layers, is observed at $2\theta = 2.81°$ ($d_{001}=3.14$ nm). In the case of organoclay modified PC, a small decrease in $2\theta$ value of the characteristic XRD peak of C15A [$2\theta = 2.79°$, ($d_{001}=3.16$ nm)] is observed along with a significant improvement in intensity of the peak which suggest for an ordered structure. In the case of PC/30PS blends, the characteristic XRD peak of the C15A is observed at $2\theta = 2.72°$ ($d_{001}=3.24$ nm), a shift of 0.09°, indicating an intercalated structure which results due the insertion of a polymer chain into the nanoclay galleries. The other XRD peak appears at $2\theta = 5.47°$ ($d_{002}=1.61$ nm) corresponding to (002) reflection. Thus the XRD results clearly suggest intercalated nanocomposite formation.

## 3.5 Thermal degradation behavior of pristine and organoclay modified PC/PS blends

Fig. 8 shows the TGA themograms of melt extruded PC, PS and PC/PS blends without and with 3% organoclay. All the pristine and compatibilized blends show two steps degradation corresponding to PS and PC phases. We have determined the degradation temperature at 10% degradation and found these to be 496°C, 451°C, 418°C, 400°C and 471°C, 456°C, 421°C, 396°C for pristine and organoclay modified PC, PC/10PS, PC/30PS and PS, respectively. It is observed that as the PS content increases degradation temperature of the blends decreases which corresponds to the lower degradation temperature of PS than PC. In addition to this, we observed a small increase in degradation temperature of organoclay modified PC/PS blends in comparison to that of the pristine blends, which attributed to the reinforcement effect of organoclay to the PC/PS blends.

## 3.6 Mechanical properties of pristine and organoclay modified PC/PS blends

The mechanical property of pristine and organoclay modified PC, PS and PC/PS blends were investigated under tensile mode (shown in Fig.9). PC shows the ductile behavior during tension. The peak in the stress-strain behavior of PC is caused by the formation of a distinct neck. This neck has propagated through the entire gauge. The specimens of PC fracture when the % elongation reaches ~60. In contrast, PS does not show any yield and fractures in brittle mode when % elongation reaches ~2.5. In the case of PC/5PS blends, upto to 5% PS content it shows ductile behavior during tension moreover its tensile strength and Young's modulus increases which could be attributed to the higher modulus of PS than PC. As the PS content increases in the blends it looses its ductility and shows brittle mode of fracture, however, its modulus keep on increasing with an increase of PS content. The organoclay modified PC/PS blends shows an improvement in the modulus of blends compared to unmodified blend but no such improvement is noted in tensile strength and toughness of the blends. The increase in the modulus of the blends may be attributed to the reinforcement effect of the organoclay to the polymer matrix. Table 1 summarizes the tensile strength, Young's modulus, elongation (in %) at break and toughness of various compositions of pristine and compatibilized PC/PS blends.

## 4. Conclusions

Melt extruded PC/PS blends and its organoclay modified nanocomposite have been studied. The FTIR, XRD and DSC results shows that pristine PC/PS blends are immiscible and only a weak interaction exist in between PC and PS phases of the blends. On the other hand, organoclay modified PC/PS blends/nanocomposite shows a compatible blend formation. The organoclay modified compatibilized PC/PS blends show a better thermal stability along with an increase in the modulus compared to the pristine PC/PS blends. These significant improvements in the properties of nanocomposite also support reinforcement effect of organoclay to the PC/PS blend matrix.

**Acknowledgements**

Mr. A.K. Singh acknowledges CSIR, India for the award of a Senior Research Fellowship. Authors acknowledge Inter University Accelerator Centre (IUAC), New Delhi for wide angle XRD measurements. We thank Professor Dhananjai Pandey for helpful discussions.

Table 1 Mechanical properties of pristine and organoclay modified PC/PS blends

| Sample | Modulus (GPa) | UTS (MPa) | Elongation (%) | Toughness MJ/m$^3$ |
|---|---|---|---|---|
| PC | 1.14 | 59.3 | 59.8 | 27.2 |
| PC/5PS | 1.15 | 61.2 | 55 | 26.6 |
| PC/10PS | 1.17 | 57.8 | 7 | 2.43 |
| PC/30PS | 1.36 | 26.5 | 2 | 0.28 |
| PS | 1.59 | 36.6 | 2.5 | 0.49 |
| PC+ 3%C15A | 1.39 | 66.8 | 8.5 | 5.3 |
| PC/10PS + 3%C15A | 1.41 | 41.1 | 3.5 | 0.79 |
| PC30PS+ 3%C15A | 1.51 | 28.3 | 2 | 0.3 |
| PS+ 3%C15A | 1.70 | 42.3 | 2.8 | 0.63 |

**Figures**

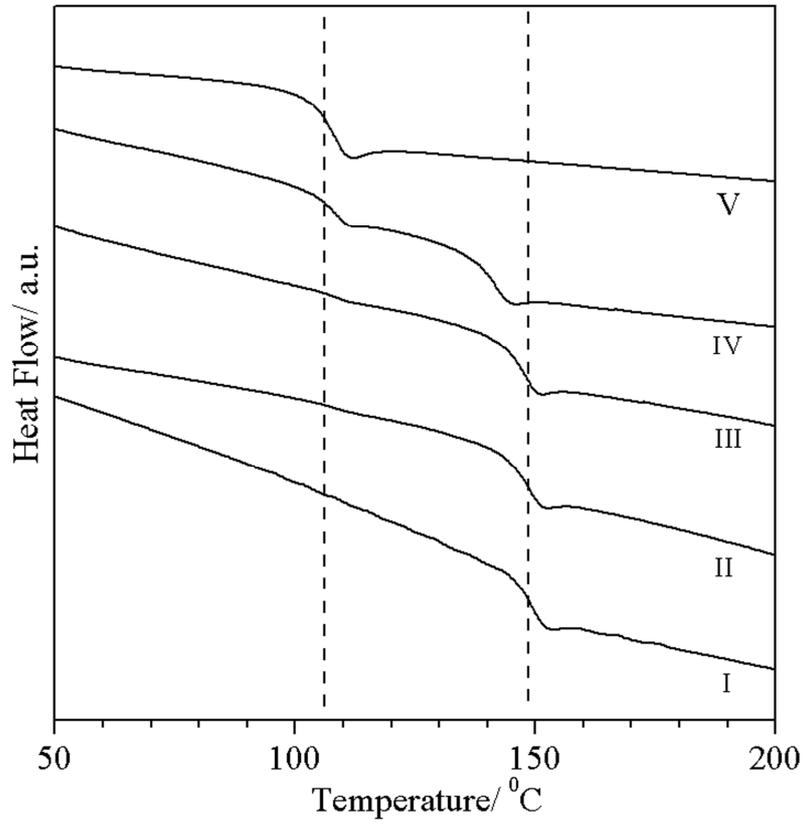

**Figure 1** Second run DSC thermograms of: (I) pure PC, (II) PC/5PS, (III) PC/10PS, (IV) PC/30PS blends, and (V) pure PS.

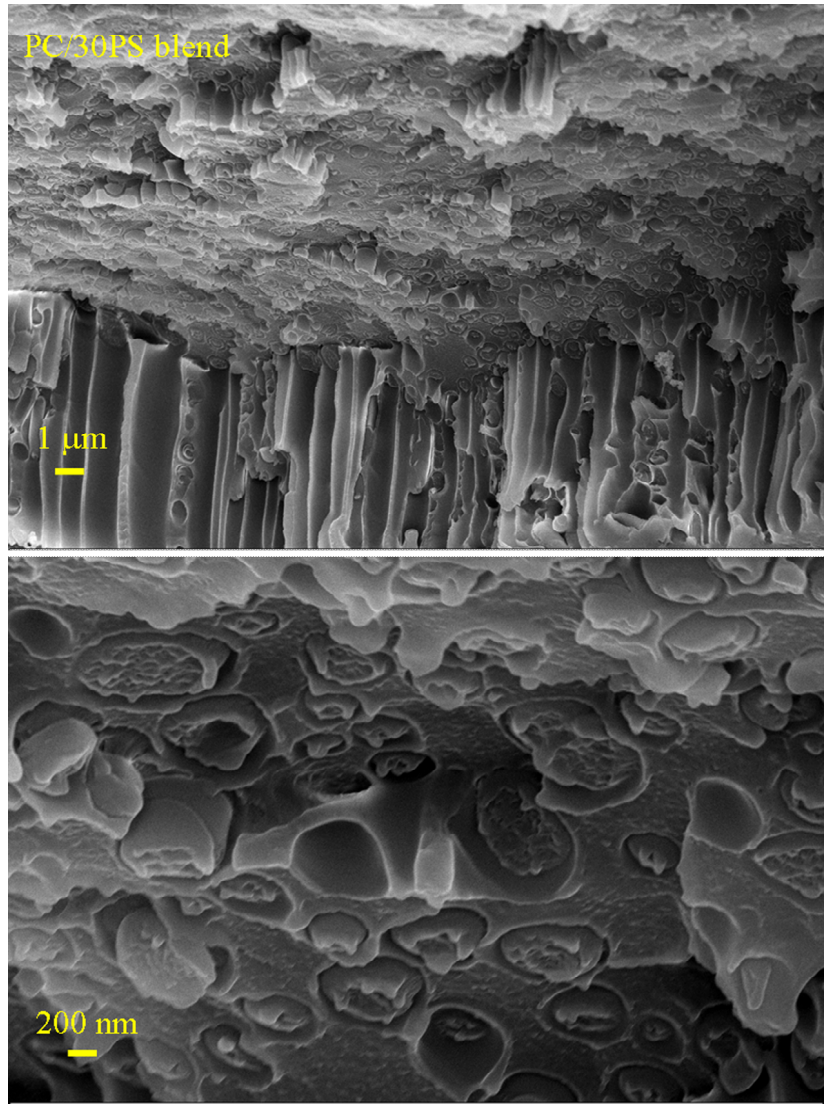

**Figure 2** SEM images of cryo-fractured surface of pristine PC/30PS blend at different magnifications.

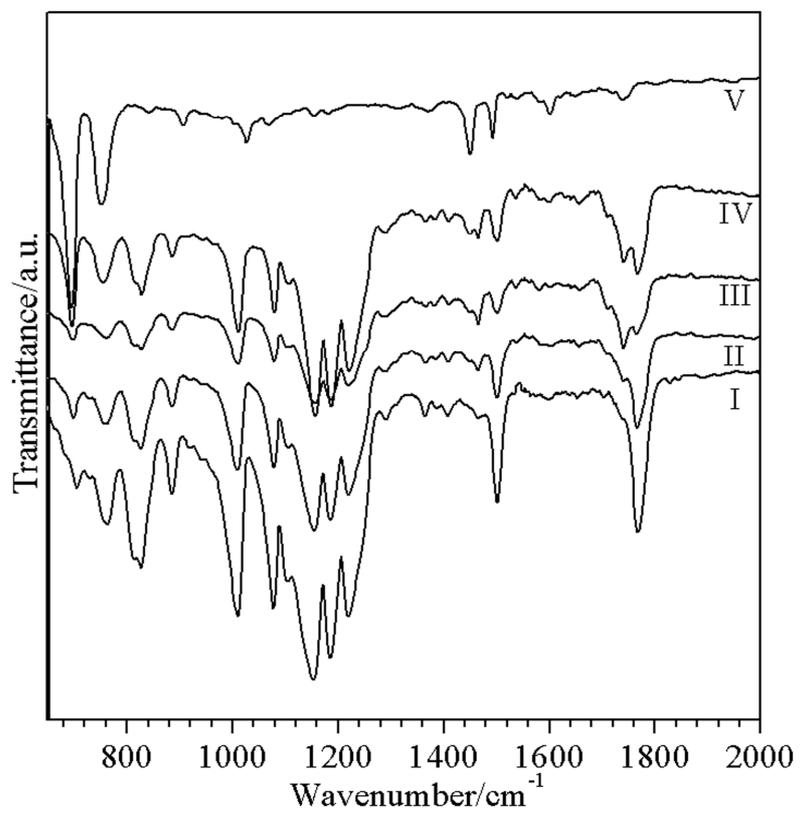

**Figure 3** FTIR spectra of: (I) pure PC, (II) PC/5PS, (III) PC/10PS, (IV) PC/30PS blends, and (V) pure PS.

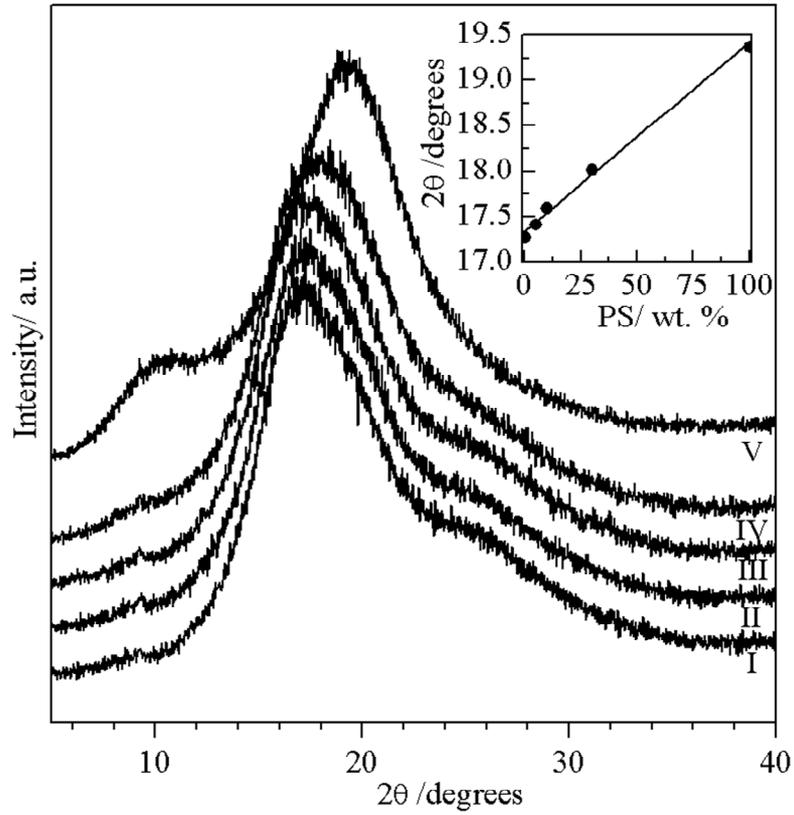

**Figure 4** XRD pattern of: (I) pure PC, (II) PC/5PS, (III) PC/10PS, (IV) PC/30PS blends, and (V) pure PS. Inset shows peak position of characteristic PC peak as a function of PS content.

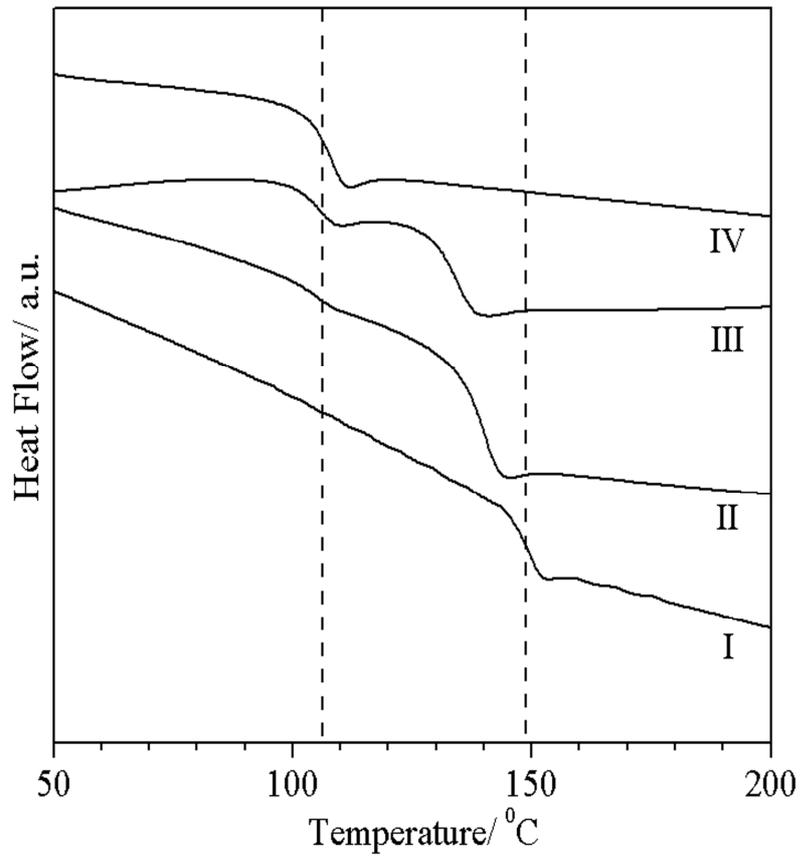

**Figure 5** DSC thermograms of: (I) pure PC, (II) PC/10PS + 3%C15A and (III) PC/30PS + 3%C15A blends, and (IV) PS.

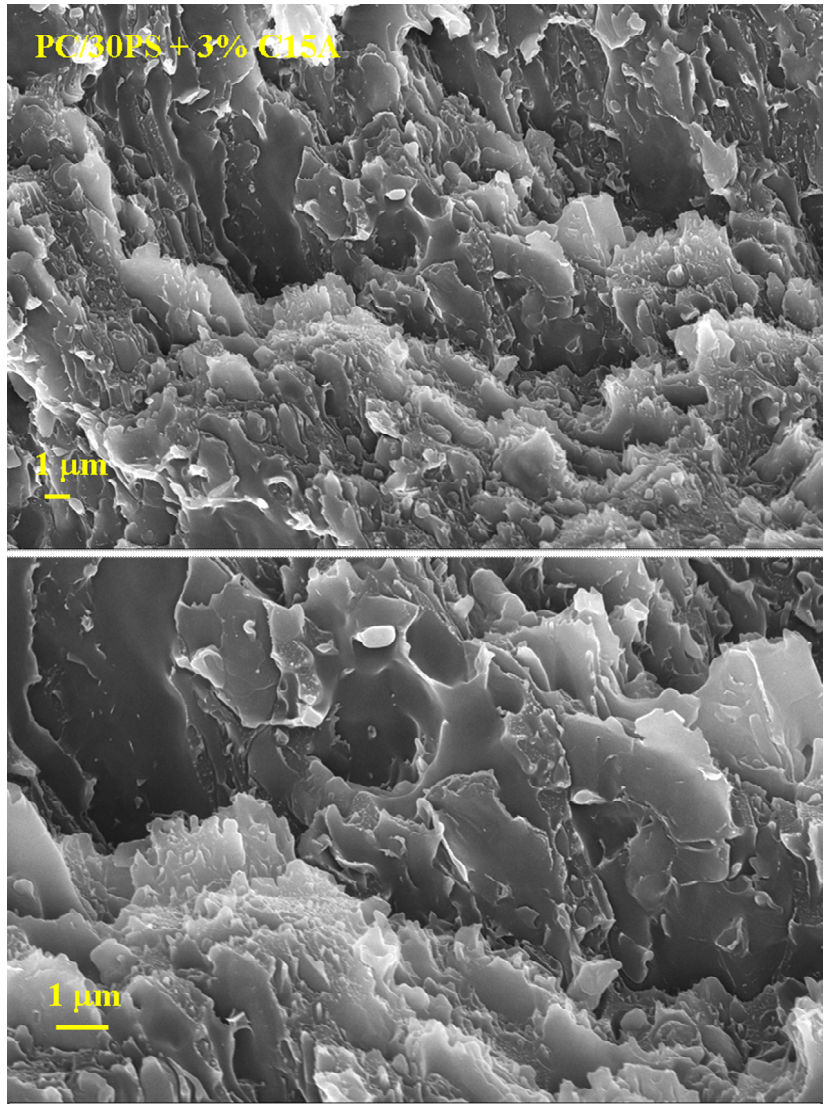

**Figure 6** SEM images of cryo-fractured surface of PC/30PS+3%C15 blend at different magnifications.

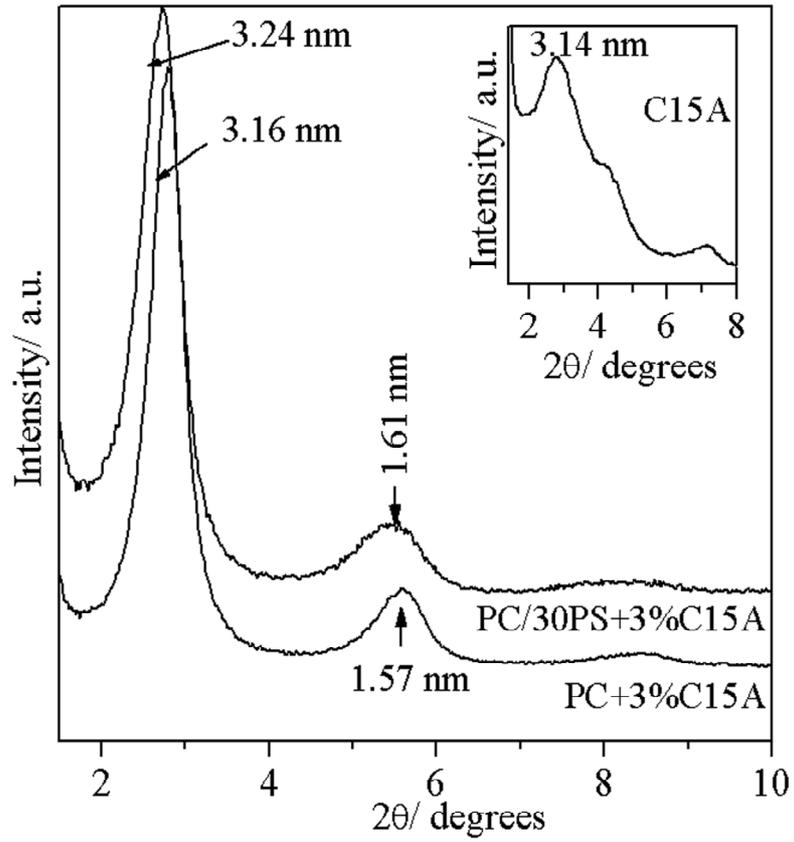

**Figure 7** XRD patterns of PC + 3%C15A and PC/30PS + 3%C15A nanocomposite. Inset shows XRD pattern of pure C15A organoclay.

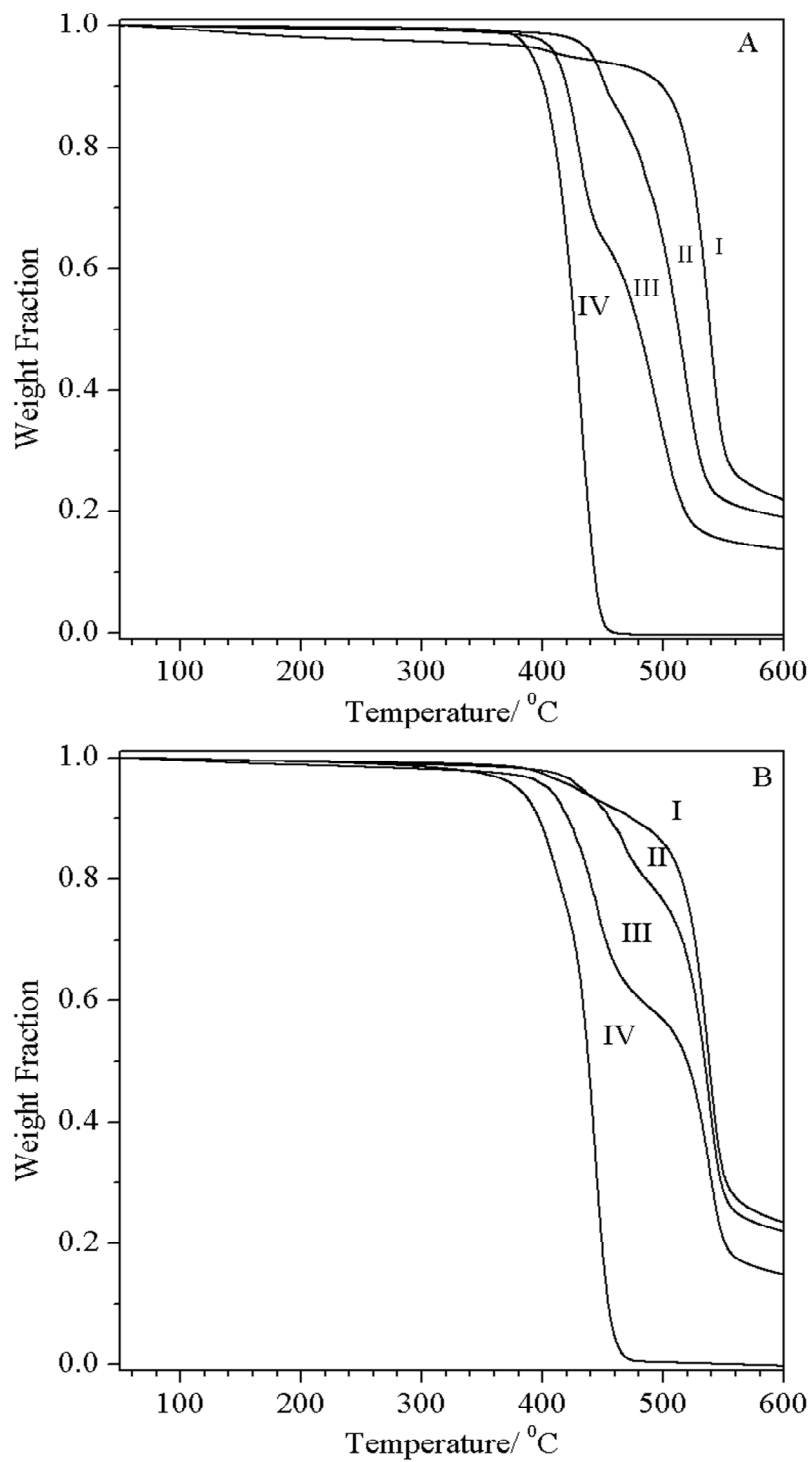

**Figure 8** TGA thermograms of: (I) pure PC, (II) PC/10PS, (IV) PC/30PS blends, and (V) PS (A) without and (B) with 3%C15A organoclay.

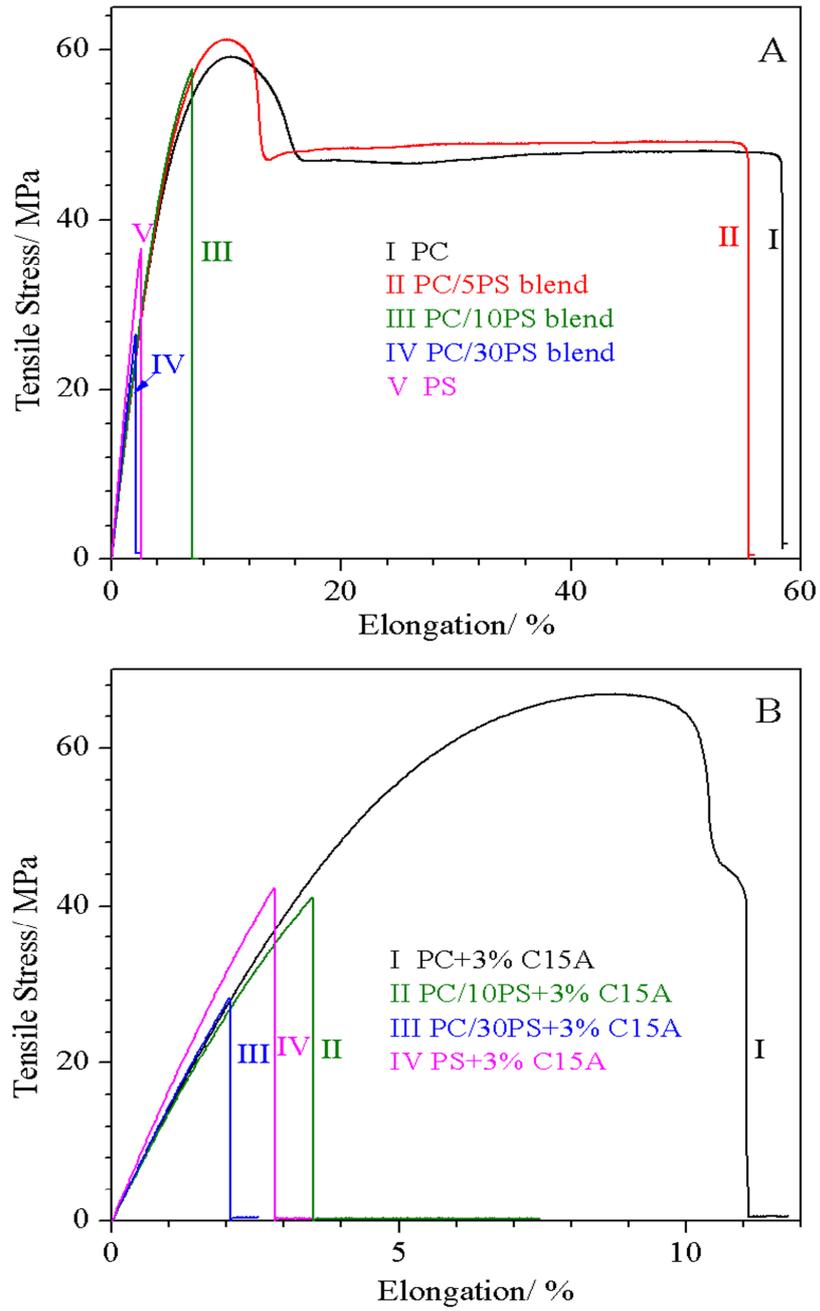

**Figure 9** Stress-strain behavior of (A) Pristine PC/PS blends, (B) PC/PS +3% C15A organoclay.